\documentclass[12pt]{article}
\pdfoutput=1

\usepackage[bulletsep]{collref}
\usepackage{amssymb,graphicx}
\usepackage[intlimits]{amsmath}
\usepackage{bbm}
\usepackage[small]{subfigure}
\usepackage{tikz}
\usepackage{MnSymbol}

\makeatletter \@addtoreset{equation}{section} \makeatother

\makeatletter
\let\old@startsection=\@startsection
\let\oldl@section=\l@section
\renewcommand{\@startsection}[6]{\old@startsection{#1}{#2}{#3}{#4}{#5}{#6\mathversion{bold}}}
\renewcommand{\l@section}[2]{\oldl@section{\mathversion{bold}#1}{#2}}
\makeatother

\makeatletter
\let\old@makecaption=\@makecaption
\def\@makecaption{\small\old@makecaption}
\makeatother

\renewcommand{\leq}{\leqslant}

\newcommand{\Q}{\mathcal{Q}}
\newcommand{\sdetr}{\mathop{\mathrm{Sdet}}\nolimits'}
\newcommand{\sdetrd}{\mathop{\mathrm{Sdet}}\nolimits_{\rm r}}

\begin{document}

\thispagestyle{empty}
\begin{flushright}\footnotesize
\texttt{NORDITA 2021-042} 
\vspace{0.6cm}
\end{flushright}

\renewcommand{\thefootnote}{\fnsymbol{footnote}}
\setcounter{footnote}{0}

\begin{center}
{\Large\textbf{\mathversion{bold} Duality Relations for Overlaps \\ of Integrable Boundary States in AdS/dCFT}
\par}

\vspace{0.8cm}

\textrm{Charlotte~Kristjansen$^{1}$, Dennis~M\"uller$^{1}$ and
Konstantin~Zarembo$^{1,2}$\footnote{Also at ITEP, Moscow, Russia}}
\vspace{4mm}

\textit{${}^1$Niels Bohr Institute, Copenhagen University, Blegdamsvej 17, 2100 Copenhagen, Denmark}\\
\textit{${}^2$Nordita, KTH Royal Institute of Technology and Stockholm University,
Hannes Alfv\'ens v\"ag 12, SE-106 91 Stockholm, Sweden}\\
\vspace{0.2cm}
\texttt{kristjan@nbi.dk, dennis.muller@nbi.ku.dk, zarembo@nordita.org}

\vspace{3mm}


\par\vspace{1cm}

\textbf{Abstract} \vspace{3mm}

\begin{minipage}{13cm} 

The  encoding  of all possible sets of Bethe equations for a spin chain with $SU(N|M)$ symmetry into a $QQ$-system calls for an  expression of spin chain overlaps entirely in terms of $Q$-functions.  We take a significant step towards deriving such a universal 
formula in the case of overlaps between Bethe eigenstates and integrable boundary states, of relevance for AdS/dCFT, by determining the transformation properties of the overlaps under fermionic as well as bosonic dualities which allows us to move between any two  descriptions of the spin chain encoded in the $QQ$-system. 
An important part of our analysis involves introducing a suitable regularization for  singular Bethe root
configurations.

\end{minipage}
\end{center}

\vspace{0.5cm}


\newpage
\setcounter{page}{1}
\renewcommand{\thefootnote}{\arabic{footnote}}
\setcounter{footnote}{0}

\section{Introduction} 
Duality relations certainly interweave the AdS/CFT correspondence as well as its various defect versions at the macroscopical level~\cite{Maldacena:1998re,Karch:2000gx}. We will address a class of duality relations which rather pertain
to the microscopical level.  These are dualities which characterize  the integrable super
spin chain underlying the AdS/CFT correspondence~\cite{Minahan:2002ve, Beisert:2003tq,Beisert:2003yb, Beisert:2005fw} and its associated integrable boundary states describing the D3-D5 defect set-up~\cite{deLeeuw:2015hxa,Buhl-Mortensen:2015gfd,Buhl-Mortensen:2017ind,Gombor:2020kgu,Komatsu:2020sup,Kristjansen:2020mhn,Gombor:2020auk}.

The spectral problem of ${\cal N}=4$ SYM as well as a number of other problems, some of which can be related to the free energy of the integrable spin chain~\cite{Gromov:2015dfa,Harmark:2017yrv, Cavaglia:2018lxi},  have been effectively dealt with exploiting these dualities that take the form of a set of so-called QQ-relations~\cite{Tsuboi:1998ne} (for reviews, see
\cite{Gromov:2017blm,Kazakov:2018ugh,Levkovich-Maslyuk:2019awk}) which  for these particular problems can be elevated to a  quantum spectral 
curve~\cite{Gromov:2013pga,Gromov:2014caa}. The QQ relations, a set of relations between certain Q-functions,
encode the various ways that one can choose the vacuum of the spin chain and the sequence of excitations
at the various levels of nesting.

Addressing correlation functions of ${\cal N}=4$ SYM amounts to studying wave functions and norms of spin chain 
eigenstates and the quantum spectral curve approach is typically not directly applicable.  A possible strategy is the separation of 
variables method~\cite{Sklyanin_1995} which have  been applied to certain sub-classes of correlation 
functions \cite{Gromov:2020fwh,Cavaglia:2021mft}.  Here we shall address the simplest possible correlation functions of ${\cal N}=4$ SYM namely one-point functions of the integrable D3-D5 probe brane set-up. These one-point
functions can be expressed as overlaps between Bethe eigenstates and integrable boundary states in the form of either
matrix product states or valence bond states~\cite{deLeeuw:2015hxa,Buhl-Mortensen:2015gfd,Kristjansen:2020mhn}.  

Closed expressions found for these overlaps~\cite{deLeeuw:2015hxa,Buhl-Mortensen:2015gfd,deLeeuw:2016umh,deLeeuw:2018mkd,Kristjansen:2020mhn} have lead to a derivation of an asymptotic all-loop formula for
the one-point functions in question~\cite{Buhl-Mortensen:2017ind,Gombor:2020kgu,Komatsu:2020sup,Kristjansen:2020mhn,Gombor:2020auk}, as well as for three-point functions involving a single trace operator and
two giant gravitons both in ${\cal N}=4$ SYM~\cite{Jiang:2019zig}  and in ABJM theory~\cite{Chen:2019kgc,Yang:2021hrl}, and  have sparked novel developments in statistical physics~\cite{Piroli:2017sei,Piroli:2018don,Piroli:2018ksf,Pozsgay:2018dzs} where the overlaps are of relevance for the study of quantum quenches. The overlap formulas contain
a universal part expressed in terms of the super determinant of the Gaudin matrix~\cite{Gaudin:1976sv} of the Bethe state in question and a dynamical factor
depending on the boundary state and being expressible as a product of Q-functions.

As a first step towards finding an expression for the one-point functions entirely in terms of Q-functions and thus making
them amenable to the quantum spectral curve approach we complete our earlier initiated study of the behavior of the super determinant of the Gaudin matrix under the duality transformations encoded in the QQ-system\footnote{At a full non-perturbative level the one-point correlator maps to a worldsheet g-function \cite{Jiang:2019xdz}, potentially calculable in any integrable field theory with boundary \cite{Dorey:2004xk}. Amenable to a combinatorial analysis \cite{Kostov:2018dmi,Kostov:2019fvw}, the g-function has recently been connected with the quantum spectral curve formalism \cite{Caetano:2020dyp}.}.  The duality transformations
consist of two classes of transformations, bosonic and fermionic. Both types correspond to exchanging the role of a vacuum
configuration and an excitation at some nesting level. A fermionic duality implies a change of statistics for certain excitations,
and corresponds to a change of Dynkin diagram for the underlying super Lie algebra, whereas a bosonic duality does not
involve any change of statistics or of Dynkin diagram. In the present paper we focus on the bosonic dualities.


Our paper is organized as follows: We begin in section~\ref{sec:singular}  by discussing known overlap formulas for the Heisenberg spin chain and complete the existing picture by extending the results to the case of singular Bethe root
configurations. While interesting in its own right this extension is also mandatory for the overarching  goal of  our work as duality transformations of non-singular Bethe root configurations are known to introduce singular roots~\cite{Kristjansen:2020vbe}. In 
section~\ref{sec:Dualities} we give a short review of the QQ-relations for integrable spin chains of $SU(N|M)$ type, illustrating the idea with the simplest possible example of the spin chain based on the super Lie algebra $SU(2|1)$. 
Subsequently, in section~\ref{sec:Bosonic}, we turn to determining the transformation properties of the super determinant of the Gaudin matrix under bosonic dualities, starting with the simplest case of the Heisenberg spin chain, moving on the $SU(2|1)$ spin chain and finishing with the most general case. Finally, section~\ref{sec:Conclusion} contains our conclusion.

\section{Integrable boundary states and overlaps \label{sec:singular}}

\subsection{Boundary states}

We start with boundary states in the Heisenberg model, which we review in quite some detail. The model is defined by the Hamiltonian
\begin{equation}
 H=\sum_{l=1}^{L}\left(1-P_{l,l+1}\right),
\end{equation} 
where $P_{l,l'}$ is the permutation operator, and is solved by Bethe ansatz 
 \begin{eqnarray} \label{BetheSU(2)}
-1 &=& \frac{Q_\theta^{-} (u_{j})}{Q_\theta^{+} (u_{j})}\,
\frac{\mathcal{Q}^{++}(u_{j})}{\mathcal{Q}^{--}(u_{j})}\equiv \,{\rm e}\,^{i\chi _j},
\end{eqnarray}
written here in terms of the Q-functions
\begin{equation}
 \mathcal{Q}(u)=\prod_{j=1}^{M}\left(u-u_j\right).
\end{equation}
The trivial Q-function 
\begin{equation}\label{Q-theta}
 Q_\theta (u)=u^L,
\end{equation}
acts as a source in the Bethe equations.
We use the standard notations for rapidity shifts:
\begin{equation}
 f^{[\pm q]}(u)=f\left(u\pm\frac{iq}{2}\right),\qquad 
 f^\pm\equiv f^{[\pm 1]},\qquad f^{++}\equiv f^{[\pm 2]}.
\end{equation}

The Bethe eigenstates $\left|\left\{u_j\right\}\right\rangle$ are $\mathfrak{su}(2)$ highest weights:
\begin{equation}\label{hw-cnd}
 S^+\left|\left\{u_j\right\}\right\rangle=0,\qquad S^3\left|\left\{u_j\right\}\right\rangle=\left(\frac{L}{2}-M\right)\left|\left\{u_j\right\}\right\rangle,
\end{equation}
with the total spin normalized as
\begin{equation}
 S^i=\sum_{l=1}^{L}\frac{\sigma _{l}^i}{2}\,.
\end{equation}
Other members of the multiplet are generated by repeated application of  $S^-$.

The commuting charges of the integrable hierarchy have definite parity under reversal of the spin chain's orientation, typically chosen to alternate with $n$: $P^{-1}Q_nP=(-1)^nQ_n$. Parity interchanges $u_j$ with $-u_j$, and an eigenvalue of $Q_n$ is an even function of Bethe roots if $n$ is even and an odd function if $n$ is odd. 

An integrable boundary state is a state annihilated by all parity-odd charges: $Q_{2n+1}\left|B\right\rangle=0$  \cite{Ghoshal:1993tm,Piroli:2017sei}.  The ensuing selection rule imposes parity invariance on the set of rapidities $\{u_j\}=\{-u_j\}$ so long as the overlap  $\left\langle B\right.\!\left| \{u_j\}\right\rangle$ is non-zero. The particle content of a boundary state thus consists of momentum-conjugate pairs $(p,-p)$. Crossing (were it well-defined) would map each pair to a single particle reflecting off a spatial boundary \cite{Ghoshal:1993tm}. The paired structure of the boundary state guarantees that reflection is elastic and proceeds without particle production. Originally proposed for relativistic systems \cite{Ghoshal:1993tm}, where crossing is well defined, the definition is extended to spin chains by  analogy  \cite{Piroli:2017sei}. 

While integrable boundary states have never been completely classified, known examples fall into two broad categories. One type of boundary states is obtained by associating to each allowed spin state, $s$,  a $k\times k$ matrix,
$\sigma^s$, and 
 tracing the product of $L$ such matrices  ($L$ is the length of the spin chain) over the $k$-dimensional auxiliary space \cite{deLeeuw:2015hxa,Piroli:2017sei}:
\begin{equation}
 \left\langle {\rm MPS}\right|=\sum_{\left\{s_l\right\}}^{}\mathop{\mathrm{tr}}
 \sigma ^{s_1}\ldots \sigma^{s_L}\left\langle s_1\ldots s_L\right|.
\end{equation}
Not any choice of matrices $\sigma ^s$ gives an integrable boundary state, but many examples where it does are known. For instance,  an MPS generated by $\sigma ^\uparrow=\sigma ^1,~~\sigma ^\downarrow=\sigma ^2$ in the Heisenberg model is integrable. 

Another common type of boundary states is built from entangled two-site blocks:
\begin{equation}\label{VBSD}
 \left\langle {\rm VBS}\right|=\left\langle D\right|^{\otimes\frac{L}{2}},\qquad 
 \left\langle D\right|=\sum_{s\,s'}^{}K_{ss'}\left\langle ss'\right|,
\end{equation}
and can be called Valence Bond States (VBS).
In the Heisenberg model any VBS is integrable, independently of the coefficients  $K_{ss'}$~\cite{Pozsgay:2018ixm}.

All known boundary states in the Heisenberg model actually descend from the generalized dimer: 
\begin{equation}\label{gendimer}
 \left\langle D_\nu \right|=(1+\nu )\left\langle \uparrow\downarrow \right|
 +(1-\nu )\left\langle \downarrow\uparrow\right|.
\end{equation}
Applying an $SU(2)$ rotation one gets:
\begin{eqnarray}\label{dimer->VBS}
 \left\langle D_\nu \right|
\,{\rm e}\,^{i\beta S^1}\,{\rm e}\,^{i\alpha S^3}
&=&(\cos\beta +\nu )\left\langle \uparrow\downarrow\right|
+(\cos\beta -\nu )\left\langle \downarrow\uparrow\right|
\nonumber \\
&&+i\,{\rm e}\,^{i\alpha }\sin\beta \left\langle \uparrow\uparrow\right|
+i\,{\rm e}\,^{-i\alpha }\sin\beta\left\langle \downarrow\downarrow\right|,
\end{eqnarray}
a generic two-site state.
One can also show that \cite{Piroli:2017sei}
\begin{equation}\label{dimer->MPS}
 \left\langle {\rm MPS}\right|=i^{-\frac{L}{2}}\left(
 \left\langle {\rm VBS}_{1}\right|+ \left\langle {\rm VBS}_{-1}\right|
 \right)\,{\rm e}\,^{\frac{i\pi S^1}{2}},
\end{equation}
where the subscripts on VBS refer to the subscripts on the associated D, cf.\ eqns.~(\ref{VBSD}) and~(\ref{gendimer})
Both types of boundary states admit higher-rank generalizations, albeit
at higher rank not all VBS are integrable. Likewise, higher-rank MPS are related to VBS by symmetry transformations only in particular cases  \cite{deLeeuw:2019ebw} but not in general. With rather few exceptions (see sec.~4.2 of \cite{deLeeuw:2019ebw}) boundary states are defined on a spin chain of even length. We shall always take $L$ to be even.

\subsection{Overlap formulas}

Integrable boundary states have a number of remarkable properties, in particular their overlaps with on-shell Bethe states admit a concise determinant representation:
\begin{equation}
 {\left\langle B\right.\!\left|\left\{u_j\right\} \right\rangle}
 \propto\prod_{j}^{}v(u_j)\sqrt{\mathop{\mathrm{Sdet}}G}.
\end{equation}
The function $v(u)$ depends on the boundary state at hand, while the determinant factor  is universal  and depends only on the
 Gaudin matrix $G$, the Jacobian of the Bethe equations (\ref{BetheSU(2)}):
\begin{equation}
 G_{jk}=\frac{\partial \chi _j}{\partial u_k}
 =\delta _{jk}\left(\frac{L}{u_j^2+\frac{1}{4}}-\sum_{m}^{}K_{jm}\right)+K_{jk},
 \qquad K_{jk}=\frac{2}{\left(u_j-u_k\right)^2+1}\,.
\end{equation}

The overlap formula in this form was first derived for the N\'eel state ($|{\rm VBS}_{1}\rangle$)  in the Heisenberg model \cite{Brockmann:2014a,Brockmann2014,Brockmann:2014b} by exploiting connections \cite{Pozsgay:2009} to the six-vertex partition function with specific boundary conditions \cite{Tsuchiya:qf}. As for higher-rank spin chains, rigorous derivations are seldom available, but many overlap formulas were conjectured \cite{deLeeuw:2016umh,deLeeuw:2018mkd,Kristjansen:2020mhn,Yang:2021hrl} or deduced from scattering theory \cite{Jiang:2019xdz,Gombor:2020kgu,Gombor:2020auk}, coordinate Bethe ansatz \cite{Jiang:2020sdw} and TBA \cite{deLeeuw:2019ebw}.

\begin{figure}[t]
\begin{center}
 \centerline{\includegraphics[width=8cm]{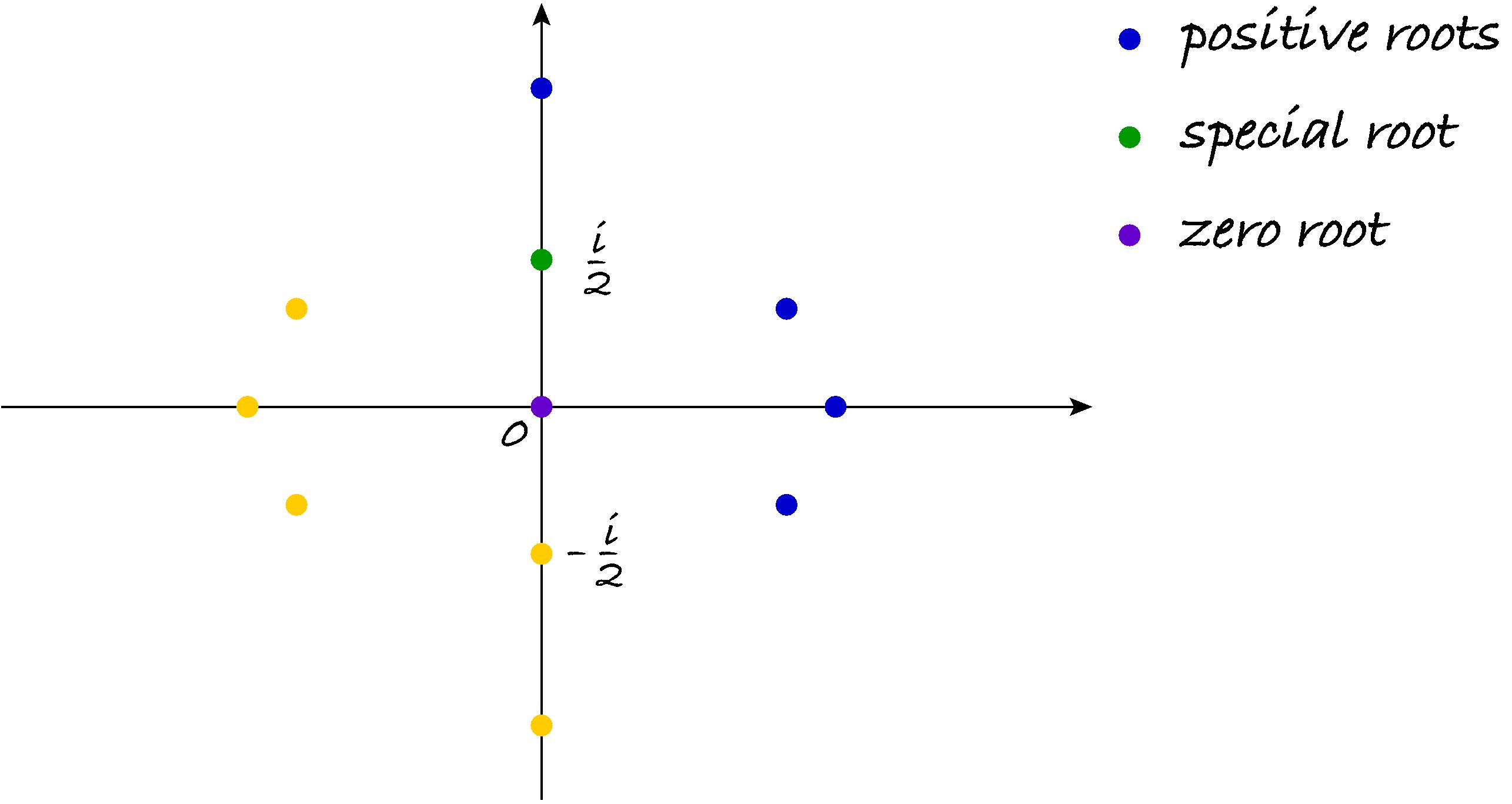}}
\caption{\label{roots-fig}\small A symmetric configuration of Bethe roots.}
\end{center}
\end{figure}

As discussed above, the Bethe roots in $\left\langle B\right.\!\left|\{u_j\} \right\rangle$ must be invariant under reflection: $\left\{u_j\right\}=\left\{-u_j\right\}$. 
A typical symmetric configuration is illustrated in fig.~\ref{roots-fig}. The roots mostly form pairs $(u_j,-u_j)$ but a solitary root at zero is also allowed. We pick one root from each pair, it does not matter which one, for instance the root with a positive real part, and call this root positive. Reflection acts as a $\mathbbm{Z}_2$ permutation (represented by a matrix $\Omega $)  on rows and columns of the Gaudin matrix,  and it is this $\mathbbm{Z}_2$ action that defines the superdeterminant:
\begin{equation}
 \mathop{\mathrm{Sdet}}G=\,{\rm e}\,^{\mathop{\mathrm{tr}}\Omega \ln G}.
\end{equation}

Additional complications arise for singular solutions of the Bethe ansatz equations with roots at $\pm i/2$  \cite{Bethe:1931hc,Avdeev:1985cx,Beisert:2003xu,Nepomechie:2013mua,Nepomechie:2014hma}. The Bethe-ansatz equations require regularization in this case. The cleanest way to proceed is via Baxter equations or the QQ-relations, not dealing with Bethe roots and circumventing the problem altogether \cite{Marboe:2016yyn}. Some entries of the Gaudin matrix become singular at $u_j=\pm i/2$, compelling us to treat singular solutions separately.

We parameterize a symmetric Bethe state in terms of positive roots $\mathbf{u}=\{u_j\}_{j=1\ldots M'}$ and two discrete parameters $\delta _z$ and $\delta _s$ indicating additional roots at zero and at $\pm i/2$:
\begin{eqnarray}
 &\left|\mathbf{u}\right\rangle=\left|\left\{u_j,-u_j\right\}\right\rangle&
 \qquad \delta _{\mathbf{z}}=0,~~\delta _{\mathbf{s}}=0
\label{state1} \\
 &\left|\mathbf{u}\right\rangle=\left|\left\{u_j,-u_j,0\right\}\right\rangle&
 \qquad \delta _{\mathbf{z}}=1,~~\delta _{\mathbf{s}}=0
\label{state2} \\
 &\left|\mathbf{u}\right\rangle=\left|\left\{u_j,-u_j,i/2,-i/2\right\}\right\rangle&
 \qquad \delta _{\mathbf{z}}=0,~~\delta _{\mathbf{s}}=1
 \label{state3} \\
  &\left|\mathbf{u}\right\rangle=\left|\left\{u_j,-u_j,0,i/2,-i/2\right\}\right\rangle&
 \qquad \delta _{\mathbf{z}}=1,~~\delta _{\mathbf{s}}=1,
\label{state4}
\end{eqnarray}
and introduce the abridged Q-function 
\begin{equation}
 Q(u)=\prod_{j=1}^{M'}\left(u^2-u^2_j\right),
\end{equation}
with the product over positive roots only. 
The total number of roots is $M=2M'+2\delta _{\mathbf{s}}+\delta _{\mathbf{z}}$, while the total momentum is
\begin{equation}
 \,{\rm e}\,^{iP}=(-1)^{\delta _{\mathbf{z}}+\delta _{\mathbf{s}}}.
\end{equation}

The master overlap formula for the $SU(2)$ spin chain can be written in the following way (see \cite{Pozsgay:2018ixm} or \cite{Gombor:2021uxz})\footnote{The phase of the overlap is not really well-defined, and depends on the phase convention for the Bethe eigenstates. Our convention corresponds to the Algebraic Bethe Ansatz as described in \cite{Faddeev:1996iy}.}:
\begin{equation}\label{master}
 \frac{\left\langle {\rm VBS}_\nu \right|\,{\rm e}\,^{i\beta S^1}\left|\mathbf{u}\right\rangle}{\left\langle \mathbf{u}\right.\!\left| \mathbf{u}\right\rangle^{\frac{1}{2}}}
 =2^{\delta _{\mathbf{s}}}i^{(L+1)\delta  _{\mathbf{z}}}\left(i\sin\beta \right)^{\frac{L}{2}-M}\,
 \frac{\mathcal{Q}\left(\frac{i\nu }{2}\right)}{\sqrt{Q(0)Q\left(\frac{i}{2}\right)}}
 \sqrt{\mathop{\mathrm{Sdet}}G}\,.
\end{equation}
The prefactor contains both normal and abridged Q-functions, the latter evaluated at the argument where the normal Q-function would be zero for special states.

This formula holds for  highest weights, for descendants one can use the identity
\begin{equation}
 \,{\rm e}\,^{i\beta S^1}=\,{\rm e}\,^{i\tan\frac{\beta }{2}\,S^-}
 \left(\cos\frac{\beta }{2}\right)^{2S^3}
 \,{\rm e}\,^{i\tan\frac{\beta }{2}\,S^+},
\end{equation}
take into account (\ref{hw-cnd}), expand in $\tan \beta /2$, and normalize by
\begin{equation}
 \left\langle \mathbf{u}\right|\left(S^+\right)^n\left(S^-\right)^n\left|\mathbf{u}\right\rangle
 =\frac{n!(L-2M)!}{(L-2M-n)!}\,\left\langle \mathbf{u}\right.\!\left|\mathbf{u} \right\rangle.
\end{equation}
Because the dependence on $\beta $ in the master formula is so simple, the $\mathfrak{su}(2)$ quantum numbers enter only through a combinatorial prefactor.

For example, the MPS overlaps following from (\ref{dimer->MPS})  are given by
\begin{equation}\label{MPS-master}
 \frac{\left\langle {\rm MPS}\right|\left(S^-\right)^{2n}\left|\mathbf{u}\right\rangle}{\left\langle \mathbf{u}\right|\left(S^+\right)^{2n}\left(S^-\right)^{2n}\left|\mathbf{u}\right\rangle^{\frac{1}{2}}}
 =2i^{-M}
\mathbbm{C}_{nML} \sqrt{\frac{Q\left(\frac{i}{2}\right)}{Q(0)}\,\mathop{\mathrm{Sdet}}G}\,,
\end{equation}
where
\begin{equation}\label{comb-fac-desc}
 \mathbbm{C}_{nML}=\frac{\left(\frac{L}{2}-M\right)!}{n!\left(\frac{L}{2}-M-n\right)!}\sqrt{\frac{(2n)!\left(L-2M-2n\right)!}{(L-2M)!}}\,,
\end{equation}
in agreement with \cite{deLeeuw:2017dkd}.
For all special states the MPS overlaps
vanish, and it is easy to understand why: states with $\delta _{\mathbf{z}}+\delta _{\mathbf{s}}=1$ are orthogonal to MPS by momentum conservation. The states with
$\delta _{\mathbf{z}}=\delta _{\mathbf{s}}=1$ carry zero momentum but contain an odd number of roots while the number of flipped spins in MPS is necessarily even.

Diagonalizing 
$\Omega $ converts
the superderminant into a ratio of ordinary determinants of rank $(M'+\delta _{\mathbf{z}}+\delta _{\mathbf{s}})\times (M'+\delta _{\mathbf{z}}+\delta _{\mathbf{s}})$ and $M'\times M'$   \cite{Brockmann:2014a}:
\begin{equation}
 \mathop{\mathrm{Sdet}}G=\frac{\det G^+}{\det G^-}\,, \label{superdet}
\end{equation}
whereas the Gaudin determinant factorizes as
\begin{equation}
 \det G=\det G^+\det G^-. \label{detfactorization}
\end{equation}
The matrix elements of the factors are
\begin{eqnarray}\label{G+-su(2)}
 G^\pm_{jk}&=&\delta _{jk}\left(\frac{L-\delta _{\mathbf{s}}}{u_j^2+\frac{1}{4}}
 -\frac{3\delta _{\mathbf{s}}}{u_j^2+\frac{9}{4}}
 -\frac{2\delta _{\mathbf{z}}}{u_j^2+1}
 -\sum_{m}^{}K_{jm}^+\right)+K^\pm_{jk}
\nonumber \\
G^+_{j\mathbf{z}}&=&\frac{2\sqrt{2}}{u_j^2+1}=G^+_{\mathbf{z}j}
\nonumber \\
G^+_{j\mathbf{s}}&=&\frac{1}{u_j^2+\frac{1}{4}}+\frac{3}{u_j^2+\frac{9}{4}}=G^+_{\mathbf{s}j}
\nonumber \\
G^+_{\mathbf{zz}}&=&4L-\frac{16\delta _{\mathbf{s}}}{3}-\sum_{m}^{}\frac{4}{u_m^2+1}
\nonumber \\
G^+_{\mathbf{zs}}&=&\frac{8\sqrt{2}}{3}=G^+_{\mathbf{sz}}
\nonumber \\
G^+_{\mathbf{ss}}&=&L-\frac{8\delta _{\mathbf{z}}}{3}-\sum_{m}^{}
\left(\frac{1}{u_m^2+\frac{1}{4}}+\frac{3}{u_m^2+\frac{9}{4}}\right),
\end{eqnarray}
where
\begin{equation}\label{Kjk}
 K^\pm_{jk}=\frac{2}{(u_j-u_k)^2+1}\pm \frac{2}{(u_j+u_k)^2+1}\,.
\end{equation}
The factorization 
formula that takes into account states with a root at zero was derived in \cite{Brockmann:2014b}.  Extension to  singular states with roots  at $\pm i/2$ is new, we believe, and can be justified by theta-regularization discussed in Appendix~\ref{thetas-appendix}.

\subsection{Nested Bethe Ansatz}

Given a Dynkin diagram for a super Lie algebra of type $SU(N|M)$ and an associated set of  Dynkin labels one
can write down a set of Bethe equations for an integrable nearest neighbor spin chain where each spin chain site carries
the representation defined by the Dynkin labels~\cite{Saleur:1999cx}. The Bethe equations can be compactly expressed in terms of 
$Q$-functions, that in turn depend on a number of Bethe roots, and read 
\begin{equation}\label{QBAEs}
(-1)^{F_a+1}= \frac{Q_\theta^{[-q_a]} (u_{aj})}{Q_\theta^{[+q_a]} (u_{aj})}\,
 \prod_{b}^{}\frac{\mathcal{Q}_{b}^{[+M_{ab}]}(u_{aj})}{\mathcal{Q}_{b}^{[-M_{ab}]}(u_{aj})}\equiv e^{i\chi_{aj}} .
\end{equation}
Here $M_{ab}$, $a,b \in \{1,\ldots,N+M-1\}$ are elements of the (tridiagonal) Cartan matrix corresponding to the chosen Dynkin
diagram and $q_a$, $a\in \{1,\ldots,N+M-1\}$ are the Dynkin labels, and $F_a$ is the fermionic parity of the $a$th node: $F_a=0$ for $M_{aa}\neq 0$; $F_a=1$ for $M_{aa}=0$. There is one $Q$-function, $\mathcal{Q}_b$, $b\in \{1,\ldots, N+M-1\}$ associated with each node in  the Dynkin diagram and in addition there is a trivial Q-function $Q_{\theta}$ defined in (\ref{Q-theta}). 
The functions $\mathcal{Q}_a$ are polynomials whose zeros are the Bethe roots:
\begin{equation}
 \mathcal{Q}_a(u)= \prod_{j=1}^{\mathcal{K}_a}(u-u_{aj}),
\end{equation}
where the $\mathcal{K}_a$ are excitation numbers that  characterize the spin chain  eigenstate.

The norm of a Bethe eigenstate is encoded in the determinant of the Gaudin matrix \cite{Gaudin:1976sv,Korepin:1982gg}
\begin{equation}
G_{aj,bk}=\frac{\partial \chi_{aj}}{\partial u_{bk}}.
\end{equation}
Bethe eigenstates that have non-trivial overlaps with integrable boundary states of VBS or MPS nature are characterized
by the Bethe roots being paired as $\{u_{aj},-u_{aj}\}$, possibly up a single root at zero\footnote{Parity $\Omega $ may act non-trivially on the nodes of the Dynkin diagram: $u_{aj}\rightarrow -u_{\sigma (a)j}$, involving a transposition $\sigma $. We will not consider this possibility, but examples can be found in \cite{Jiang:2019xdz,Yang:2021hrl}.}. For such root configurations the
determinant of the Gaudin matrix factorizes as in eqn.~(\ref{detfactorization})
and the overlaps are expressed in terms of the super determinant of $G$, defined in eqn.~(\ref{superdet}),
in combination with a number of $\mathcal{Q}$-functions. 

The factors are given by the obvious generalization of (\ref{G+-su(2)})\footnote{The markers $\delta _{a\mathbf{z}}$, $\delta _{a\mathbf{s}}$ indicate if the node $a$ carries zero and special roots (special roots sit at $\pm iq_a/2$). By convention, $1/q_a$, $1/M_{ab}$ are understood to be zero if $q_a$, $M_{ab}$ are zero.}:
\begin{eqnarray}
 G_{aj,bk}^\pm&=&\delta _{ab}\delta _{jk}\left\{\frac{Lq_a}{u_{aj}^2+\frac{q_a^2}{4}}
-\sum_{c}^{}\left[
\frac{\delta _{c\mathbf{s}}\left(M_{ac}+q_c\right)}{u_{aj}^2+\frac{\left(M_{ac}+q_c\right)^2}{4}}+\frac{\delta _{c\mathbf{s}}\left(M_{ac}-q_c\right)}{u_{aj}^2+\frac{\left(M_{ac}-q_c\right)^2}{4}}+\frac{\delta _{c\mathbf{z}}M_{ac}}{u_{aj}^2+\frac{M_{ac}^2}{4}}
\right]
\right.
\nonumber \\
&&\left.\vphantom{\sum_{c}^{}\left[
\frac{\delta _{c\mathbf{s}}\left(M_{ac}+q_c\right)}{u_{aj}^2+\frac{\left(M_{ac}+q_c\right)^2}{4}}+\frac{\delta _{c\mathbf{s}}\left(M_{ac}-q_c\right)}{u_{aj}^2+\frac{\left(M_{ac}-q_c\right)^2}{4}}+\frac{\delta _{c\mathbf{z}}M_{ac}}{u_{aj}^2+\frac{M_{ac}^2}{4}}
\right]}
-\sum_{cl}^{}K_{aj,cl}^+
\right\}+K_{aj,bk}^\pm
\nonumber \\
G_{aj,b\mathbf{z}}^+&=&\frac{\sqrt{2}\,M_{ab}}{u_{aj}^2+\frac{M_{ab}^2}{4}}=G_{b\mathbf{z},aj}^+
\nonumber \\
G_{aj,b\mathbf{s}}^+&=&\frac{M_{ab}+q_b}{u_{aj}^2+\frac{\left(M_{ab}+q_b\right)^2}{4}}
+\frac{M_{ab}-q_b}{u_{aj}^2+\frac{\left(M_{ab}-q_b\right)^2}{4}}
=G_{b\mathbf{s},aj}^+
\nonumber \\
G_{a\mathbf{z},b\mathbf{z}}^+&=&
\delta _{ab}\left[\frac{4L}{q_a}-\sum_{c}^{}
\left(\frac{8\delta _{c\mathbf{s}}M_{ac}}{M_{ac}^2-q_c^2}+\frac{4\delta _{c\mathbf{z}}}{M_{ac}}\right)
-\sum_{cl}^{}\frac{2M_{ac}}{u_{cl}^2+\frac{M_{ac}^2}{4}}
\right]+\frac{4}{M_{ab}}
\nonumber \\
G_{a\mathbf{z},b\mathbf{s}}^+&=&\frac{4\sqrt{2}\,M_{ab}}{M_{ab}^2-q_b^2}
=G_{b\mathbf{s},a\mathbf{z}}^+
\nonumber \\
G_{a\mathbf{s},b\mathbf{s}}^+&=&\delta _{ab}\left\{
\frac{L}{q_a}-\sum_{c}^{}\left[
\frac{4\delta _{c\mathbf{s}}M_{ac}}{M_{ac}^2-\left(q_a-q_c\right)^2}
+\frac{4\delta _{c\mathbf{s}}M_{ac}}{M_{ac}^2-\left(q_a+q_c\right)^2}
+\frac{4\delta _{c\mathbf{z}}M_{ac}}{M_{ac}^2-q_a^2}
\right]
\vphantom{\sum_{cl}^{}\left[
\frac{M_{ac}+q_a}{u_{cl}^2+\frac{\left(M_{ac}+q_a\right)^2}{4}}
+\frac{M_{ac}-q_a}{u_{cl}^2+\frac{\left(M_{ac}-q_a\right)^2}{4}}
\right]}
\right.
\nonumber \\
&&\left.
-\sum_{cl}^{}\left[
\frac{M_{ac}+q_a}{u_{cl}^2+\frac{\left(M_{ac}+q_a\right)^2}{4}}
+\frac{M_{ac}-q_a}{u_{cl}^2+\frac{\left(M_{ac}-q_a\right)^2}{4}}
\right]
\right\}
\nonumber \\
&&
\vphantom{\sum_{cl}^{}\left[
\frac{M_{ac}+q_a}{u_{cl}^2+\frac{\left(M_{ac}+q_a\right)^2}{4}}
+\frac{M_{ac}-q_a}{u_{cl}^2+\frac{\left(M_{ac}-q_a\right)^2}{4}}
\right]}
+\frac{4M_{ab}}{M_{ab}^2-(q_a-q_b)^2}+\frac{4M_{ab}}{M_{ab}^2-(q_a+q_b)^2}
\end{eqnarray}
with
\begin{equation}
 K^\pm_{aj,bk}=\frac{M_{ab}}{\left(u_aj-u_{bk}\right)^2+\frac{M_{ab}^2}{4}}\pm
 \frac{M_{ab}}{\left(u_{aj}+u_{bk}\right)^2+\frac{M_{ab}^2}{4}}\,.
\end{equation}

The Gaudin superdeterminant will be the main focus of our study.

\section{Duality Transformations \label{sec:Dualities}}

For simplicity, let us specialize to the simplest possible super Lie algebra $SU(2|1)$ and let us choose the Cartan matrix and the Dynkin labels to be 
 \begin{equation} \label{Cartan-Dynkin}
M=\begin{bmatrix}
 2  & -1  \\ 
  -1 & 0 
  \end{bmatrix}, \hspace{0.7cm}
  q=\begin{bmatrix}
  1 \\ 
  0 \\ 
 \end{bmatrix},
\end{equation}
which in particular implies picking the Dynkin diagram $\bigcircle$\!\! ---\!\! $\bigotimes$. 
The Bethe equations then take the following form
 \begin{eqnarray}
-1 &=& \frac{Q_\theta^{-} (u_{1j})}{Q_\theta^{+} (u_{1j})}\,
\frac{\mathcal{Q}_{1}^{++}(u_{1j})}{\mathcal{Q}_{1}^{--}(u_{1j})} \frac{\mathcal{Q}_{2}^{-}(u_{1j})}{\mathcal{Q}_{2}^{+}(u_{1j})},
\label{Bethe1}\\
 1&=& \frac{\mathcal{Q}_{1}^{-}(u_{2j})}{\mathcal{Q}_{1}^{+}(u_{2j})}\,.\label{Bethe2}
\end{eqnarray}
We note that the roots associated with
the fermionic node, $\{u_{2j}\}$, do not have any self-interactions. Fermionic duality relies precisely on this property of the Bethe equations and amounts to a change of variables  where the fermionic roots $\{u_{2j}\}_{j=1}^{\mathcal{K}_2}$ are eliminated in favor of a set of dual roots $\{\tilde{u}_{2k}\}_{k=1}^{\widetilde{\mathcal{K}}_2}$ which
are the roots of the polynomial  $\widetilde{Q}_2$ defined via the relation
\begin{equation}\label{duality-Q}
\mathcal{Q}_{1}^{+}(u)-\mathcal{Q}_{1}^{-}(u) =\mathcal{Q}_2(u) \widetilde{\mathcal{Q}}_2(u).
\end{equation}
We note that the degree of the polynomial $\widetilde{\mathcal{Q}}_2$ is given by
\begin{equation}
\widetilde{\mathcal{K}}_2= \mathcal{K}_1-\mathcal{K}_2-1\label{duality-1}.
\end{equation}

After the duality transformation  the Bethe equations take the form
\begin{eqnarray}
1 &=& \frac{{Q}_\theta^{-} (u_{1j})}{{Q}_\theta^{+} (u_{1j})}\,
 \frac{\mathcal{\widetilde{Q}}_{2}^{-}(u_{1j})}{\mathcal{\widetilde{Q}}_{2}^{+}(u_{1j})},\label{DualBethe1}\\
 1&=& \frac{\mathcal{Q}_{1}^{-}(\tilde{u}_{2j})}{\mathcal{Q}_{1}^{+}(\tilde{u}_{2j})}.\label{DualBethe2}
\end{eqnarray}
This simple example illustrates the general phenomenon that a fermionic duality transformation after a given node leaves the Bethe equations associated with that node unchanged but changes 
the nature of the roots at the neighboring nodes from bosonic to fermionic or vice versa, thus effectively changing the Dynkin
diagram underlying the description.
Furthermore, in the case where a non-zero Dynkin 
label is associated with the fermionic node, the Dynkin label of the node and  its neighboring nodes will change (see~\cite{Kristjansen:2020vbe} for further discussion focussed on overlap formulas). 
Due to the tri-diagonal nature of the
Cartan matrix, in general only the Bethe equations corresponding to the node itself and its two neighbors can be affected by the duality transformation. Similarly, only the part of the Gaudin matrix that refers to the phases $\chi_{aj}$  of the dualized node, $a$, and its two neighbors, $a-1$ and $a+1$, will change. 

Under a general fermionic duality transformation the Gaudin super determinant transforms in an 
astonishingly simple way. More precisely, starting from a regular root configuration with all roots paired and dualizing after the  node $a$ one finds the
following transformation formula~\cite{Kristjansen:2020vbe}
\begin{equation}\label{transformation_formula}
S\!\det \tilde{G}\,\propto\, \frac{Q_a(0) \widetilde{Q}_a(0)}{Q_{a+1}(\frac{i}{2}) Q_{a-1}(\frac{i}{2})} \,S\! \det G,
\end{equation}
where the constant of proportionality depends on the excitation numbers as well as the local Dynkin labels and Cartan matrix elements. 
In the $Q$-functions appearing in~(\ref{transformation_formula}) the roots at $\pm i/2$ and at zero are left out. 
The relation holds semi-off-shell meaning that the Bethe roots $u_{aj}$, $\tilde{u}_{aj}$ must fulfill a duality relation \`{a} la eqn.\
(\ref{duality-Q}) but the remaining roots can be chosen arbitrarily. With this transformation rule for the super determinant the
overlaps between Bethe eigenstates of the $PSU(2,2|4)$ spin chain and the integrable boundary states of relevance for domain wall versions of ${\cal N}=4$ SYM  transform covariantly~\cite{Kristjansen:2020vbe}. We note that by means of fermionic duality transformations we can move between all the possible Dynkin diagrams of the underlying super Lie algebra. 

The fermionic dualities only constitute a sub-set of the possible duality transformations of the Bethe equations. The various 
possible forms of the Bethe equations reflect the different possible choices of vacuum state and of excitations at the various
levels of nesting.  All duality transformations are conveniently expressed in terms of a set of $QQ$-relations which in turn can  be encoded in a Hasse diagram~\cite{Tsuboi:1998ne}, see also~\cite{Gromov:2017blm,Kazakov:2018ugh,Levkovich-Maslyuk:2019awk}.
 For a Lie-algebra of type $SU(N|M)$ there is a total of $2^{M+N}$ $Q$-functions and each of them is 
associated with a node in the corresponding  Hasse diagram.

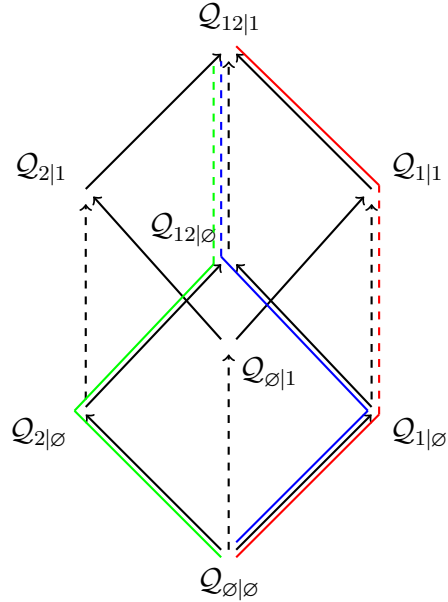
\begin{figure}[htb]
\begin{center}
  \begin{tikzpicture}
        \draw (0,0) node[anchor=north]{$\mathcal{Q}_{\emptyset|\emptyset}$};
        \draw (2,2) node[anchor=north west] {$\mathcal{Q}_{1|\emptyset}$};
        \draw (-2,2) node[anchor=north east] {$\mathcal{Q}_{2|\emptyset}$};
        \draw (0,2.8) node[anchor=north west] {$\mathcal{Q}_{\emptyset|1}$};
        \draw (0,4) node[anchor=south east] {$\mathcal{Q}_{12|\emptyset}$};
        \draw (2,4.8) node[anchor=south west] {$\mathcal{Q}_{1|1}$};
        \draw (-2,4.8) node[anchor=south east] {$\mathcal{Q}_{2|1}$};
        \draw (0,6.8) node[anchor=south] {$\mathcal{Q}_{12|1}$};
        \draw[thick,->] (0.1,0.1)--(1.9,1.9);
        \draw[thick,->] (-0.1,0.1)--(-1.9,1.9);
        \draw[thick,->,dashed] (0,0.1)--(0,2.7);
        \draw[thick,->,dashed] (1.9,2.1)--(1.9,4.7);
        \draw[thick,->,dashed] (-1.9,2.1)--(-1.9,4.7);
        \draw[thick,->] (-1.9,2)--(-0.1,3.9);
        \draw[thick,->] (1.9,2)--(0.1,3.9);
        \draw[thick,->,dashed] (0,4.1)--(0,6.6);
        \draw[thick,->] (0.1,2.9)--(1.8,4.8);
        \draw[thick,->] (-0.1,2.9)--(-1.8,4.8);
        \draw[thick,->] (1.9,4.9)--(0.1,6.7);
        \draw[thick,->] (-1.9,4.9)--(-0.1,6.7);
               \draw[thick,blue] (0.1,0.2)--(1.85,1.95);
          \draw[thick,blue] (1.85,1.95)--(-0.1,4);
         \draw[thick,blue,dashed] (-0.1,4)--(-0.1,6.6);
        \draw[thick,red] (0.1,0)--(2,1.9);
        \draw[thick,red,dashed] (2,1.9)--(2,4.95);
        \draw[thick,red] (2,4.95)--(0.1,6.8);
            \draw[thick,green] (-0.1,0.)--(-2.05,1.95);
               \draw[thick,green] (-2.05,1.95)--(-0.2,3.9);
       \draw[thick,green,dashed] (-0.2,3.9)--(-0.2,6.6);
    \end{tikzpicture}

\end{center}
\caption{The Hasse diagram of $SU(2|1)$. Each path from the bottom to the top node corresponds to a particular
form of the Bethe equations. The blue path turns into the red path under a fermionic duality transformation and 
into the green path under a bosonic duality transformation.\label{Hasse1}}
\end{figure}

Again, for simplicity, let us specialize to $SU(2|1)$ for which we show the Hasse diagram in~figure~\ref{Hasse1}. Each path from the bottom node to the top node corresponds to a particular version of the Bethe equations and a fermionic duality transformation of the Bethe equations corresponds to flipping two segments of the path vertically across a plaquette. 

Our original Bethe equations~(\ref{Bethe1}) and (\ref{Bethe2}) correspond to the blue path whereas the Bethe equations
after dualization, i.e.\ eqns.\ (\ref{DualBethe1}) and (\ref{DualBethe2}) correspond the red path with the following identification
of $\mathcal{Q}$-functions
\begin{eqnarray}
&& \mathcal{Q}_{\emptyset|\emptyset}=Q_{\theta}, \hspace{1.cm}  \mathcal{Q}_{1|\emptyset}=\mathcal{Q}_{1}, \hspace{1.0cm}  \mathcal{Q}_{12|\emptyset}=\mathcal{Q}_2,\\
&& \mathcal{Q}_{12|1}=1,\hspace{1.25cm} \mathcal{Q}_{1|1}=\widetilde{\mathcal{Q}}_2.
\end{eqnarray}
In addition to the fermionic
duality already considered one can perform a bosonic duality transformation on the first node on the Dynkin diagram 
\mbox{$\bigcircle$\!\! ---\!\! $\bigotimes$}. This amounts to a change of variables from the roots $u_{1j}$ to a set of dual roots 
$\tilde{u}_{1k}$, where the dual variables are the roots of the polynomial $\widetilde{\mathcal{Q}}_1$, which is to be identified with $\mathcal{Q}_{2|\emptyset}$, and which is defined via
\begin{equation}\label{withmomentum}
\mathcal{Q}_1^+(u)\widetilde{\mathcal{Q}}_1^-(u)-\mathcal{Q}_1^-(u)\widetilde{\mathcal{Q}}_1^+(u) = Q_{\theta} (u)\mathcal{Q}_2(u). 
\end{equation}
A bosonic duality transformation corresponds to flipping the path in the Hasse diagram horizontally across a plaquette and
in the present case transforms the blue path into the green path. 
Under this change of variables the Bethe equations turn into
 \begin{eqnarray}
-1 &=& \frac{Q_\theta^{-} (\tilde{u}_{1j})}{Q_\theta^{+} (\tilde{u}_{1j})}\,
\frac{\mathcal{\widetilde{Q}}_{1}^{++}(\tilde{u}_{1j})}{\mathcal{\widetilde{Q}}_{1}^{--}(\tilde{u}_{1j})} \frac{\mathcal{Q}_{2}^{-}(\tilde{u}_{1j})}{\mathcal{Q}_{2}^{+}(\tilde{u}_{1j})},
\label{Bethe1dual}\\
 1&=& \frac{\mathcal{\widetilde{Q}}_{1}^{-}(u_{2j})}{\mathcal{\widetilde{Q}}_{1}^{+}(u_{2j})}. \label{Bethe2dual}
\end{eqnarray}
This simple example illustrates the fact that a bosonic duality transformation as opposed to a fermionic one does not change the nature of the Bethe equations and thus does not change the Dynkin diagram. Likewise the Dynkin labels on all nodes  are invariant under the transformation.  Finally, as before, due to the tri-diagonal form of the Cartan matrix only the part of the Gaudin matrix that refers to the phases $\chi_{aj}$  of the dualized node, $a$, and its two neighbors, $a-1$ and $a+1$, will change. 

In the following we will investigate the transformation properties of the super determinant of the Gaudin matrix under bosonic dualities with the aim of determining  whether a bosonic equivalent of the fermionic transformation formula~(\ref{transformation_formula}) exists.  We will start by considering the simplest possible case of $SU(2)$, where the
bosonic node is necessarily momentum carrying, from there proceed to $SU(2|1)$, where the bosonic mode can be either momentum carrying or not, and  finally consider the case of the most general three-node Dynkin diagram having a bosonic node in the middle. 
As we shall see, the transformation rule is the same in all three cases and it is natural to expect that it holds for any Dynkin diagram of  $\mathfrak{su}(N|M)$ type.  

\section{Bosonic Dualities \label{sec:Bosonic}}

In this section we study the behavior of the super determinant of the Gaudin matrix under bosonic dualities. We start from
the $SU(2)$ spin chain and from there move on to more general cases.

\subsection{$SU(2)$: Q-functions and regularization}

The Bethe equations of the $SU(2)$ Heisenberg spin chain are given by eqn. (\ref{BetheSU(2)}) which 
we note constitute a special case of the Bethe equations~(\ref{Bethe1}) and (\ref{Bethe2}) with $\mathcal{Q}_2=1$ (and thus
no roots of type ${u_{2j}}$). Accordingly, the bosonic duality transformation is given by  eqn.~(\ref{withmomentum}). With
the appropriate simplifications after a slight change in normalization the latter becomes:
\begin{equation}
 \Q^+\widetilde{\Q}^--\Q^-\widetilde{\Q}^+=i(2{M}-L-1)u^L. \label{SU(2)QQbare}
\end{equation}

The roots of $\widetilde{\Q}$ lie "beyond the equator" \cite{Pronko:1998xa}: the initial number of excitations $M$ is limited by the highest-weight condition (\ref{hw-cnd}) to be $M\leq L/2$, while after dualizing the number of roots becomes
\begin{equation}
{\widetilde{M}}=L-{M}+1,
\end{equation}
and exceeds $L/2$. 

The dual roots are ambiguous:  a linear combination $\widetilde{\Q}+\alpha \Q$ solves the QQ-equation equally well as $\widetilde{\Q}$. This could have caused unnecessary complications, but Bethe eigenstates of interest are reflection invariant and that saves the day. Assuming all roots are paired, the original Baxter polynomial is an even function. It is then natural to require the dual Q-function to have an opposite parity:
\begin{equation}
 \Q(-u)=\Q(u),\qquad \widetilde{\Q}(-u)=-\widetilde{\Q}(u).
\end{equation}
This choice makes the "large" solution unique.

It is true in general that
\begin{equation}
 \widetilde{\delta }_{\mathbf{z}}=1-\delta _{\mathbf{z}},
\end{equation}
and, somewhat less obviously, that
\begin{equation}
 \widetilde{\delta }_{\mathbf{s}}=1-\delta _{\mathbf{s}}.
\end{equation}
In other words, if the state $\left|\{u_j\}\right\rangle$ is not special then $\{\widetilde{u}_j\}$ contain roots at $\pm i/2$, and vice versa. Same for the root at zero.

Degeneracy of the Gaudin matrix poses a more serious problem: it always happens that $\det \widetilde{G}_+=0$\footnote{We have no analytic proof of this statement, but have checked on a huge number of examples that $\det G_+=0$ so long as $M>L/2$.}. The dual superdeterminant is consequently ill defined and has to be regularized to make any meaningful statement. The regularization we use singles out the root at zero\footnote{Other possibilities include turning on the twist \cite{Gombor:2021uxz}, removing the zero eigenvector from $\widetilde{G}_+$  \cite{Cavaglia:2021mft} or dropping the special roots. We did not find any simple dualization formula in the first two cases, the last one is discussed in appendix~\ref{AltReg} and leads to results very similar to regularization adopted in the main text.}: consider an $(\widetilde{M}'+\widetilde{\delta }_{\mathbf{s}})\times (\widetilde{M}'+\widetilde{\delta }_{\mathbf{s}})$ matrix $\widetilde{G}_+'$ given by (\ref{G+-su(2)}) with $u_j$ replaced by $\widetilde{u}_j$ and with the $\widetilde{G}^+_{\mathbf{z}j}$ row and $\widetilde{G}^+_{j\mathbf{z}}$ column dropped. We then define
\begin{equation}
 \sdetr\widetilde{G}=\frac{\det \widetilde{G}_+'}{\det \widetilde{G}_-}\,.
\end{equation}

With this regularization we find the following relation between the original and the dual superdeterminant 
\begin{align}\label{sdet->sdet'}
 \sdetr \widetilde{G}= 2^{2-4\delta _{\mathbf{s}}} \mathbbm{A}_{L/2-M} \, \frac{Q(0) \widetilde{Q}(i/2)}{Q(i/2) \widetilde{Q}(0)} \,
 \mathop{\mathrm{Sdet}}G,
\end{align}
where we introduced a shorthand notation for the combinatorial prefactor:
\begin{equation}\label{A_n}
 \mathbbm{A}_{n}=\frac{(2^{n}n!)^4}{(2n)!(2n+1)!}\,.
\end{equation}

This relation, which was found by numerical investigations, has several features in common  with the transformation law found for fermionic 
dualities~\cite{Kristjansen:2020vbe}.  Incidentally, it involves the same ratio of Baxter polynomials that appears in the overlap of the ${\rm VBS}_0$ with the on-shell Bethe states. The latter is thus duality-invariant having precisely the same form in terms of original and dual roots, up to a somewhat involved pre-factor that resembles the one appearing in the overlaps with descendants, cf.~\cite{deLeeuw:2017dkd} or eq.~(\ref{comb-fac-desc}). The overlaps with ${\rm VBS}_0$ are precisely those that appear in the Bethe ansatz description of the one-point functions in the simplest domain-wall dCFT \cite{Kristjansen:2020mhn}.

\subsection{From $SU(2)$ to $SU(2|1)$}\label{2node}
As the next to simplest case, let us move on to  $SU(2|1)$. We pick the Cartan matrix and the Dynkin labels
as in the example treated in section~\ref{sec:Dualities}, i.e.\ as given in eqn.~(\ref{Cartan-Dynkin}). Hence, we consider the Dynkin diagram $\bigcircle\!\!-\!\!\!-\!\!\!-\!\!\bigotimes$ and take the bosonic node to be momentum carrying. The bosonic duality transformation corresponds  to moving from the blue path to the green path in figure~\ref{Hasse1} and 
amounts to invoking the QQ-relation given in eqn.~(\ref{withmomentum}). The number of bosonic roots before and after
the dualization are related by
\begin{equation}
\widetilde{\mathcal{K}}_1=L+\mathcal{K}_2-\mathcal{K}_1+1.
\end{equation}
For simplicity, let us consider the case where the number of original bosonic roots, $\mathcal{K}_1$, as well as  the number of  fermionic roots, $\mathcal{K}_2$, are even. Then there will be an odd number of dual bosonic roots, $\widetilde{\mathcal{K}}_1$, and hence a single dual root at zero.  
 Under these assumptions the QQ-relation is adapted to
\begin{equation}
\mathcal{Q}_1^+\widetilde{\Q}_1^--\mathcal{Q}_1^-\widetilde{\Q}_1^+ = i(2\mathcal{K}_1-L-\mathcal{K}_2-1)u^L \mathcal{Q}_2.
\end{equation}

Using the same regularization procedure as above we find from numerical investigations a transformation formula for the super determinant which generalizes the $SU(2)$ result to the nested Bethe ansatz, namely 
\begin{align}
\sdetr\widetilde{G}=  2^{2-4\delta _{1\mathbf{s}}}
\mathbbm{A}_{(L+\mathcal{K}_2)/2-\mathcal{K}_1}\, \frac{{Q}_1(0) \widetilde{Q}_1(i/2)}{{Q}_1(i/2) \widetilde{Q}_1(0)}\, \mathop{\mathrm{Sdet}}G.
\end{align}
The effect of the extra fermionic node is to replace $L $ by $ L+\mathcal{K}_2$. The same formula applies, without any change, for
$SU(3)$ where the auxiliary node is bosonic:  $\bigcircle\!\!-\!\!\!-\!\!\!-\!\!\bigcircle$. Regularization by omitting roots at $\pm i/2$ leads to a formula with the Jacobian inverted, the precise transformation rule is spelled out in the appendix~\ref{AltReg}.

\subsection{General momentum carrying case}\label{3node}

Let us consider the general case of a three-node Dynkin diagram with a momentum carrying bosonic node in the middle,
more precisely
\begin{equation}
 M=\begin{bmatrix}
 \eta_1  & -1 & 0  \\ 
  -1 &  2 & -1 \\ 
    0 & -1 & \eta_2 \\ 
 \end{bmatrix}, \hspace{0.7cm}  q=\begin{bmatrix}
 0\\
  1 \\ 
  0 \\ 
 \end{bmatrix},\hspace{0.7cm} \eta_1,\eta_2\in \{0,2\}.
\end{equation}
 In this case the relevant QQ-relation reads
\begin{equation}
\mathcal{Q}_m^+\widetilde{\mathcal{Q}}_m^--\mathcal{Q}_m^-\widetilde{\mathcal{Q}}_m^+
=i(2\mathcal{K}_m-L-\mathcal{K}_l-\mathcal{K}_r-1) u^L \mathcal{Q}_l\, \mathcal{Q}_r,
\end{equation}
where  $l,r$ and $m$ stand for left, right  and middle. We assume that the original roots are all
paired, then as before there will be a single zero root at the middle node after the dualization.  
We find a transformation formula which generalizes the two cases considered before, namely
\begin{align}
\sdetr\widetilde{G}= 2^{2-4\delta _{m\mathbf{s}}}\mathbbm{A}_{(L+\mathcal{K}_r+\mathcal{K}_l)/2-\mathcal{K}_m}\,\frac{{Q}_m(0) \widetilde{Q}_m(i/2)}{{Q}_m(i/2) \widetilde{Q}_m(0)} \mathop{\mathrm{Sdet}}G.
\end{align}
The neighboring nodes have resulted in $L$ being replaced by $L+\mathcal{K}_l+\mathcal{K}_r$ but otherwise play no role. In particular, the result is independent of whether these nodes are bosonic or fermionic. The transformation rule for the alternative regularization is displayed in the appendix~\ref{AltReg}.

\subsection{Dualizing a  non-momentum carrying node}
\begin{figure}[htb]
\begin{center}
    \begin{tikzpicture}
        \draw (0,0) node[anchor=north]{$Q_{\emptyset|\emptyset}$};
        \draw (2,2) node[anchor=north west] {$Q_{1|\emptyset}$};
        \draw (-2,2) node[anchor=north east] {$Q_{2|\emptyset}$};
        \draw (0,2.8) node[anchor=north west] {$Q_{\emptyset|1}$};
        \draw (0,4) node[anchor=south east] {$Q_{12|\emptyset}$};
        \draw (2,4.8) node[anchor=south west] {$Q_{1|1}$};
        \draw (-2,4.8) node[anchor=south east] {$Q_{2|1}$};
        \draw (0,6.8) node[anchor=south] {$Q_{12|1}$};
        \draw[thick,->] (0.1,0.1)--(1.9,1.9);
        \draw[thick,->] (-0.1,0.1)--(-1.9,1.9);
        \draw[thick,->,dashed] (0,0.1)--(0,2.7);
        \draw[thick,->,dashed] (1.9,2.1)--(1.9,4.7);
        \draw[thick,->,dashed] (-1.9,2.1)--(-1.9,4.7);
        \draw[thick,->] (-1.9,2)--(-0.1,3.9);
        \draw[thick,->] (1.9,2)--(0.1,3.9);
        \draw[thick,->,dashed] (0,4.1)--(0,6.6);
        \draw[thick,->] (0.1,2.9)--(1.8,4.8);
        \draw[thick,->] (-0.1,2.9)--(-1.8,4.8);
        \draw[thick,->] (1.9,4.9)--(0.1,6.7);
        \draw[thick,->] (-1.9,4.9)--(-0.1,6.7);
                \draw[thick,red] (-0.1,2.75)--(-2.05,4.9);
        \draw[thick,red] (-2.05,4.9)--(-0.2,6.7);
        \draw[thick,red,dashed] (-0.1,0.2)--(-0.1,2.75);        
       \draw[thick,blue] (1.75,4.9)--(0.1,6.55);
      \draw[thick,blue] (0.1,3.05)--(1.75,4.9);   
       \draw[thick,blue,dashed] (0.1,0.2)--(0.1,3.1);
    \end{tikzpicture}
\end{center}
\caption{The QQ-relation~(\ref{nomomentum}) encodes the change of path from blue to red on the Hasse diagram.\label{Hasse2}}
\end{figure}
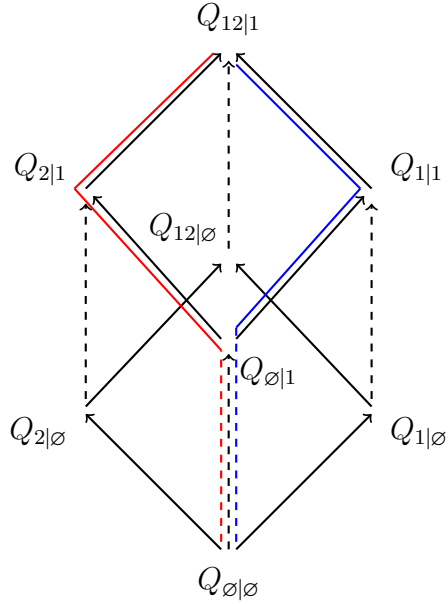

Let us go back to $SU(2|1)$ but consider a different grading with the Dynkin diagram $\bigotimes\!\!-\!\!\!-\!\!\!-\!\!\bigcircle$ where now the
bosonic node does not carry any momentum, more precisely
\begin{equation}
M=\begin{bmatrix}
 0  & -1  \\ 
  -1 & 2 
  \end{bmatrix}, \hspace{0.7cm}
  q=\begin{bmatrix}
  1 \\ 
  0 \\ 
 \end{bmatrix}. 
 \end{equation}
 The associated Bethe equations read
  \begin{eqnarray}
1 &=& \frac{Q_\theta^{-} (u_{1j})}{Q_\theta^{+} (u_{1j})}\,
 \frac{\mathcal{Q}_{2}^{-}(u_{1j})}{\mathcal{Q}_{2}^{+}(u_{1j})},
\\
 -1&=& \frac{\mathcal{Q}_{1}^{++}(u_{2j})}{\mathcal{Q}_{1}^{--}(u_{2j})}\frac{\mathcal{Q}_{1}^{-}(u_{2j})}{\mathcal{Q}_{1}^{+}(u_{2j})},
\end{eqnarray}
and correspond to the blue path on the Hasse diagram in figure~\ref{Hasse2} with the identifications
\begin{eqnarray}
&& \mathcal{Q}_{\emptyset|\emptyset}=Q_{\theta}, \hspace{1.cm}  \mathcal{Q}_{\emptyset|1}=\mathcal{Q}_{1}, \hspace{1.0cm}  \mathcal{Q}_{1|1}=\mathcal{Q}_2, \hspace{1.0cm}
 \mathcal{Q}_{12|1}=1.
\end{eqnarray}
 The bosonic duality relation  reads
 \begin{equation}
\mathcal{Q}_2^+\widetilde{\mathcal{Q}}_2^--\mathcal{Q}_2^-\widetilde{\mathcal{Q}}_2^+ = i(2\mathcal{K}_2-\mathcal{K}_1-1) \mathcal{Q}_1, \label{nomomentum}
\end{equation}
and the implied change of variables corresponds to flipping two links of the blue path horizontally across a plaquette to arrive
at the red path with the understanding that $\widetilde{\mathcal{Q}}_2$ should be identified with $Q_{2|1}$.
Assuming again the original roots to be paired, we delete the row and
the column corresponding to the zero dual root. This leads to a closed transformation formula which reads
\begin{equation}
\sdetr\widetilde{G}=  \mathbbm{A}_{\mathcal{K}_1/2-\mathcal{K}_2}\, \frac{\widetilde{Q}_2(i/2) {{Q}}_2(0)}{\widetilde{Q}_2(0) {{Q}}_2(i/2)} \mathop{\mathrm{Sdet}}G.
\end{equation}
The relative factor of $2^{2-4\delta _{\mathbf{s}}}$ is missing because neither $\Q_2$ nor $\widetilde{\Q}_2$ have special roots. For the same reason we cannot use the alternative regularization. The roots  at $\pm i/2$ cannot appear on the auxiliary node because there is no momentum term in the Bethe equations to compensate for the pole in the scattering amplitude.

One can repeat the experiment for a non-momentum carrying node in  $SU(3)$,   i.e.\ considering the Dynkin diagram $\bigcircle \!\!-\!\!\!-\!\!\bigcircle$ and
\begin{equation}
M=\begin{bmatrix}
 2  & -1  \\ 
  -1 & 2 
  \end{bmatrix}, \hspace{0.7cm}
  q=\begin{bmatrix}
  1 \\ 
  0 \\ 
 \end{bmatrix},
 \end{equation}
where one dualizes after the second node. The QQ-relation takes the same form as~(\ref{nomomentum}) and the transformation formula is again the same.

\section{Conclusion \label{sec:Conclusion}}

We have found that a combination of Q-functions and the super determinant of the Gaudin matrix for an integrable spin chain of $SU(N|M)$ type transforms covariantly under the bosonic dualities encoded in the QQ-relations of the spin chain. Such combinations are characteristic of overlaps between Bethe eigenstates and integrable boundary states, which appear in
 the study of one-point functions in AdS/dCFT.  We have earlier reported  on similar observations concerning the fermionic duality relations and pointed out that the overlap formulas of relevance  for the D3-D5 brane induced domain wall version of ${\cal N}=4$ SYM were singled out by transforming covariantly under  fermionic dualities~\cite{Kristjansen:2020vbe}. The overlaps do not have equally appealing transformation properties under bosonic dualities, except when restricted to the scalar $SO(6)$ sector. With the present work we have analyzed all 
available duality relations and it is our hope that 
our findings  could provide an important clue towards a completely universal formula for overlaps in terms of $Q$-functions thereby making the one-point function problem amenable to the quantum spectral curve approach and allowing one to pursue these observables  beyond the asymptotic regime which has so far only been possible for protected operators~\cite{Komatsu:2020sup}. The full non-perturbative answer is expected to take the form of a g-function  \cite{Dorey:2004xk}, with the Gaudin matrix superseded by a ratio of Fredholm determinants \cite{Jiang:2019xdz}. It would be interesting to understand how those transform under the symmetries of the Quantum Spectral Curve.

\subsection*{Acknowledgements}

We would like to thank Dinh-Long Vu for comments on the draft.
The work of CK and DM was supported by DFF-FNU through grant number DFF-4002-00037. The work of KZ was supported by the grant "Exact Results in Gauge and String Theories" from the Knut and Alice Wallenberg foundation and by RFBR grant 18-01-00460 A. 

\appendix

\section{Theta regularization}\label{thetas-appendix}

The Algebraic Bethe ansatz admits impurities. If those are paired: $\left\{\theta _l,-\theta _l\right\}$, $l=1\ldots L/2$, reflection symmetry is preserved and one can identify integrable boundary states as those preserving parity-odd charges of the hierarchy. 

Impurities present, the Bethe ansatz equations become
\begin{equation}
 \prod_{l=1}^{\frac{L}{2}}\frac{\left(u_j+\frac{i}{2}\right)^2-\theta _l^2}{\left(u_j-\frac{i}{2}\right)^2-\theta _l^2}=
 \prod_{k\neq j}^{}\frac{u_j-u_k+i}{u_j-u_k-i}\,.
\end{equation}
The Gaudin matrix and its factors change accordingly:
\begin{eqnarray}\label{G-theta}
 G^\pm_{jk}&=&\left\{\sum_{l=1}^{\frac{L}{2}}\left[\frac{1}{\left(u_j-\theta _l\right)^2+\frac{1}{4}}+\frac{1}{\left(u_j+\theta _l\right)^2+\frac{1}{4}}\right]
  -\frac{2\delta _{\mathbf{z}}}{u_j^2+1}-\sum_{m}^{}K^+_{jm}\right\}\delta _{jk}+K^\pm_{jk}
\nonumber \\
{G}^+_{j\mathbf{z}}&=&\frac{2\sqrt{2}}{u_j^2+1}=G^+_{\mathbf{z}j}
\nonumber \\
{G}^+_{\mathbf{zz}}&=&\sum_{l=1}^{\frac{L}{2}}\frac{2}{\theta _l^2+\frac{1}{4}}-\sum_{m}^{}\frac{4}{u_m^2+1}\,
\end{eqnarray}
with the same $K^\pm_{jk}$ as in (\ref{Kjk}).

The singular states are removed or, better say, regularized by the impurities. If $\theta _l$ are very small, the solution for singular roots is
\begin{equation}\label{epsilon-reg}
 u_{1,2}=\pm\frac{i}{2}\pm\varepsilon 
\end{equation}
with 
\begin{equation}
 \varepsilon\approx  i\prod_{k>2}^{}\frac{u_k-\frac{3i}{3}}{u_k+\frac{i}{2}}\,
 \prod_{l=1}^{\frac{L}{2}}\theta _l^2. 
\end{equation}

In the limit $\theta _l\rightarrow 0$ two apparent singularities occur in the first equation in (\ref{G-theta}). The diagonal matrix elements  $G_{11}^\pm$ corresponding to the special root, those that arise from the kinetic term in the Bethe equations, hit the pole. In addition, the scattering elements $K^\pm_{11}$ diverge as $\pm 1/[(i/2+i/2)^2+1]$. This divergence only affects $G^-_{11}$, while in $G_{11}^+$ it cancels. The divergence in the kinetic term is actually fictitious, because\footnote{Here we are taking into account that $\varepsilon \ll\theta $.}:
\begin{equation}
 \lim_{\theta \rightarrow 0}\left[
 \frac{1}{\left(\frac{i}{2}-\theta \right)^2+\frac{1}{4}}
 +\frac{1}{\left(\frac{i}{2}+\theta \right)^2+\frac{1}{4}}
 \right]=2.
\end{equation}
As a consequence, $G_{11}^+$ remains finite in the limit:
\begin{equation}
 \lim_{\theta _l\rightarrow 0}G_{11}^+
 =L-\frac{8\delta _{\mathbf{z}}}{3}-\sum_{m}^{}
\left(\frac{1}{u_m^2+\frac{1}{4}}+\frac{3}{u_m^2+\frac{9}{4}}\right).
\end{equation}

The scattering singularity in $G^-_{11}$, on the contrary, is not fake:
\begin{equation}
 G^-_{11}=\frac{i}{\varepsilon }+{\rm finite},
\end{equation}
and $\det G^-$ diverges as
\begin{equation}
 \det G^-=\frac{i}{\varepsilon }\,\det G^-_{\rm reg},
\end{equation}
where $G^-_{{\rm reg}}$ is the matrix with the first row and first column dropped. The divergence is compensated by the zero of the Q-function:
\begin{equation}
 Q\left({i}/{2}\right)=-i\varepsilon Q_{\rm reg}\left({i}/{2}\right),
\end{equation}
the product staying finite in the limit:
\begin{equation}
 \lim_{\theta _l\rightarrow 0}Q(i/2)\det G^-=Q_{\rm reg}(i/2)\det G^-_{\rm reg}.
\end{equation}

This leads to the regularization prescription in the main text. An extra factor of $2^{\delta _{\mathbf{s}}}$ in (\ref{master}) arises from removing the $i/2$ root from $Q(0)$:
\begin{equation}
 \lim_{\theta _l\rightarrow 0}Q(0)=\frac{1}{4}\,Q_{\rm reg}(0).
\end{equation}

\section{Alternative regularization}\label{AltReg}

In the main text we regularized the dual Gaudin matrix by dropping the row and column of the zero root. One can alternatively remove the roots at $\pm i/2$. This regularization works if the initial state was not special, the singular roots are then necessarily present in the dual Baxter function. 
Shifting the singular dual roots away from $\pm \frac{i}{2}$  according to (\ref{epsilon-reg}) yields a dual Gaudin matrix $\widetilde{G}$ with $\varepsilon$-poles in the upper left $2\times2$ corner but with otherwise finite matrix elements,
and we define the reduced Gaudin matrix as the matrix which is obtained by crossing out the first two rows and columns in $\widetilde{G}$:
\begin{align}
\widetilde{G}^{\text{red}}=\widetilde{G}_{ij}\Bigr|_{i,j=3,..,L-M+1} \, .
\end{align}
Since there are no divergences in $\widetilde{G}^{\text{red}}$, all Bethe roots and in particular $u_1$ and $u_2$ can directly be set to their face value.  The reduced Gaudin matrix enjoys parity invariance and can be factorized much in the same way as the Gaudin matrix without the singular roots. We define the reduced Gaudin super determinant as the ration of factors:
\begin{align}
\sdetrd\widetilde{G}=\frac{\det \widetilde{G}^{\rm red}_+}{\det\widetilde{G}^{\rm red}_-}
\end{align}

By experimenting with numerics we found that the reduced superdeterminant follows a simple transformation law:
\begin{align}
 \sdetrd \widetilde{G}=  2\mathbbm{A}_{L/2-M}\, \frac{Q(i/2) \widetilde{Q}(0)}{Q(0) \widetilde{Q}(i/2)} \,
 \mathop{\mathrm{Sdet}}G,
\end{align}
with the numerical prefactor defined in (\ref{A_n}) and   the Jacobian inverted compared to
 (\ref{sdet->sdet'}).
Interestingly, the overlaps which transform covariantly in this case are those between the Bethe eigenstates and the N\'{e}el state, or the MPS overlaps which prominently feature in the $\mathcal{N}=4$ dCFT \cite{deLeeuw:2015hxa,Buhl-Mortensen:2015gfd}.

Adding extra nodes to the Dynkin diagram only changes the numerical prefactor in the transformation law. For the general case considered in sec.~\ref{3node} the transformation formula is
\begin{align}
\sdetrd\widetilde{G}= 2\mathbbm{A}_{(L+\mathcal{K}_l+\mathcal{K}_r)/2-\mathcal{K}_m} \, \frac{{Q}_m(i/2) \widetilde{Q}_m(0)}{{Q}_m(0) \widetilde{Q}_m(i/2)} \mathop{\mathrm{Sdet}}G.
\end{align}
The two-node case from sec.~\ref{2node} is obtained by setting $\mathcal{K}_l=0$.

\bibliographystyle{nb}

\end{document}